\documentclass[a4paper,11pt]{article}

\usepackage{url} %?
\usepackage{times} %?
\usepackage{verbatim}
\usepackage[T1]{fontenc}
\usepackage[finnish, english]{babel}
\usepackage{amsmath}
\usepackage{amssymb}
\usepackage[dvipdf]{graphicx}
\usepackage{color}
\usepackage{url}
\usepackage{epstopdf}

\usepackage{ctable}
\usepackage{epsfig}
\usepackage{subfigure}
\usepackage{float}
\usepackage[small]{caption}
\usepackage{subfig}

\def\reals{\mathbb{R}}
\def\naturals{\mathbb{N}}
\def\logit{\mathop{\rm logit}\nolimits}

\def\bin{\mathop{\rm Bin}}

\newtheorem{theorem}{Theorem}

\newtheorem{remark}[theorem]{Remark}

\title{Systematic and non-systematic mortality risk in pension portfolios}

\author{Helena Aro\footnote{Department of Mathematics and Systems Analysis, Faculty of Information and Natural Sciences, Aalto University, helena.aro@aalto.fi}
\footnote{The author gratefully acknowledges the valuable advice of Prof. Teemu Pennanen of King's College, London, as well as the insightful comments of an anonymous referee.}
}

\date{}

\begin{document}

\maketitle

\begin{abstract}
We study the effects of non-systematic and systematic mortality risks on the required initial capital in a pension plan, in the presence of financial risks. We discover that for a pension plan with few members the impact of pooling on the required capital per person is strong, but non-systematic risk diminishes rapidly as the number of members increases. Systematic mortality risk, on the other hand, is a significant source of risk is a pension portfolio.

\end{abstract}

 \vspace{4mm}

\section{Introduction}

The mortality risk of a population can be decomposed into two components: systematic and non-systematic or idiosyncratic mortality risk. Systematic mortality risk refers to the uncertainty in the future development in the survival probabilities of the population. This risk is undiversifiable, and does not depend on the size of the population. However, even if future survival probabilities were known, future numbers of survivors would still be random. These fluctuations account for the non-systematic mortality risk, which diminishes as the size of the population increases, and is theoretically diversifiable by pooling. 

This paper aims to assess the effects of non-systematic and systematic mortality risks on the capital requirement of a pension plan. We consider a simple asset-liability model of a defined-benefit pension plan, and compute the required initial capital per person for varying numbers of members in the scheme. We show that for pension plans with few members the impact of pooling on the capital requirement per participant is strong, but the effect diminishes rapidly as the pool size increases.

The role of non-systematic risk in pension plans has been studied in earlier literature.
 Coppola at al. \cite{Coppola2000} consider the contributions of mortality and investment risks to the variability in the present value of liabilities, given annuity portfolios of different sizes. Olivieri \cite{Olivieri2001} also considers the impact of systematic and random fluctuations on the present value of future benefit payments under a deterministic financial structure.  Milevsky et al. \cite{Milevsky2006} show how the standard deviation of payoffs per policy diminishes to a constant as the number of policies increases.They discovered that when there are dozens of policies, the contribution of non-systematic risk is still notable, but for portfolios larger than a thousand members it reduces to negligible.  Hári et al. \cite{Hari2008} have examined the impact of non-systematic risk on a capital reserve, described as a proportion of the present value of the liabilities, required to reduce the probability of underfunding to an acceptable level. Donnelly \cite{Donnelly} considers the role of non-systematic risk in a pension plan by studying how the coefficient of variation for the liabilities of the scheme varies with the number of participants in the scheme. According to her findings, non-systematic risk in a pension portfolio decreases steeply, and becomes negligible for pension portfolios with a few hundred members.

In this paper, we show how the least amount of initial capital required to cover the liabilities of pension portfolio varies with the size of the portfolio. This choice is in line with the fact that the Solvency II directive focuses on the capital requirements of insurance companies. In addition, it is a tangible and comprehensible concept. Whereas traditional actuarial methods determine capital requirements by discounting expected cash flows, we adopt a slightly different approach. 
We consider a multi-period model of stochastic asset returns and liabilities, and determine the minimum initial capital needed to cover the liabilities  in terms of a convex risk measure, given a degree of risk aversion. Our results corroborate earlier findings, showing that as the size of the pension portfolio grows, the effect of non-systematic risk first drops sharply, then diminishes more slowly, ultimately reaching negligible levels.

Section 2 quantifies mortality risk and defines the capital requirement problem of a pension fund. Section 3 presents the numerical results. Section 4 concludes.

\section{Valuation of defined-benefit pension liabilities}

Consider a defined-benefit pension plan,
% with $\sum_{x \in X} E_{x,0}$ members. 
where the number of members aged $x$ at time $t$ is denoted by $E_{x,t}$. The number of survivors $E_{x+1,t+1}$ among the $E_{x,t}$ individuals during year $[t,t+1)$ can be described by the binomial distribution:
\begin{equation}\label{eq:bin}
E_{x+1,t+1} \sim\bin(E_{x,t},p_{x,t}),
\end{equation}
where $p_{x,t}$ is the probability that an $x$ year-old individual randomly selected at the beginning of year $t$ survives until $t+1$.

The future values of $E_{x+1,t+1}$ are obtained by sampling from $\bin(E_{x,t},p_{x,t})$. 
The uncertainty in the future values of $p_{x,t}$ represents the \emph{systematic mortality risk}. Even if the 'true' survival probabilities were known, future population sizes would still be random, which accounts for the \emph{non-systematic mortality risk}.  However, as the population grows, the fraction $E_{x+1,t+1}/[E_{x,t}p_{x,t}]$ converges in distribution to constant $1$. In large enough pools the main uncertainty comes from unpredictable variations in the future values of $p_{x,t}$, and the population dynamics are well described by $E_{x+1,t+1}=E_{x,t}p_{x,t}$. 

We assume that each alive member receives an index-linked annual benefit at times $t=1,2,\ldots,T$, until termination of the scheme at $t=T$. The yearly pension claims amount to 
\[
c_t=\frac{I_t}{I_0}\sum_{x \in X}d_{x}E_{x,t},
\]
where $I_t$ is the index value, $X \subset \naturals$ is the set of age groups in the pension plan, and the constant $d_{x}$ depends on the value of the index and accrued pension benefit at time $t=0$. We will look for the least amount of capital $w_0$ that suffices to cover the liabilities until the termination of the scheme. 

Currently the most widely-used practice for valuation of insurance liabilities is based on the \emph{actuarial 'best estimate'}, obtained as the expected value of discounted claims. This is the valuation approach also applied in Solvency II. When using this method one assumes that the portfolio is large enough to warrant not
taking into account the role of nonsystematic risk. On the other hand, Föllmer and Knispel \cite{FollmerKnispel} note that if one takes the sum of i.i.d claims, and defines the capital requirement by the \emph{entropic risk measure}  $\rho$ defined for random variables $X$ as
 %\[
% \mathcal{A}=\{ X \in L^0 \ | \ \rho(X) \le 0) \}=\{ X \in L^0 \ | \frac{1}{\gamma} \log E[e^{-\gamma x}] \le 0 \}
% \]
  \begin{equation}\label{eq:entropic}
\rho(X)=\frac{1}{\gamma} \log E[e^{-\gamma X}],
\end{equation}
then the required capital per individual does not diminish as $n$ increases. More precisely, they show that if
$X_1, X_2, \cdots, X_n$ are i.i.d. random variables on a probability space $(\Omega, \mathcal{F}, \mathbb{P})$ and 
$S_n=X_1+ X_2+ \cdots +X_n$, then $\rho(S_n)=n\rho(X)$. 

Neither of the above valuation approaches considers underlying systematic risk factors affecting the liability cash flows, such as uncertainty in the joint survival probabilities of the entire population. Moreover, the sufficiency of capital also depends on how it is invested in the financial markets. The 'best estimate' approach corresponds with the assumption that all of the capital is invested in fixed-income instruments, which does not comply with the investment policy of a typical pension insurer. On the other hand, the setting of Föllmer and Knispel essentially assumes that the wealth is stored in a cash account. 

In order to study the effects of non-systematic and systematic risks on capital requirements, we apply the valuation approach described in \cite{cashflow}. 
At each $t$, the insurer pays out $c_t$ and invests the remaining wealth $w_t$ in financial markets.
The investment returns are modelled as a stochastic process, which is dependent on the chosen investment strategy used by the insurer. 
As in  \cite{cashflow} we define the value of liabilities as the least initial capital that enables the investor to hedge the liability cash flows with given risk tolerance.

The liabilities $(c_t)_{t=0}^T$ and returns $(R_t)_{t=0}^T$ are modelled as adapted stochastic processes, and problem can be formulated as
\begin{equation}
\begin{aligned}
& {\text{min}}& &w_0 \ \ \text{over} \ w \in \mathcal{N} \\
& \text{subject to}& & w_t \leq R_tw_{t-1} - c_t\ \ t=1,\ldots,T \\
%&&&h_{t,j} \in \mathcal{D}_t, t=1,\ldots,T \\
%&&&h_{t,j} \ge 0 \ \ j \in J \setminus \{0\}  \\
&&& \rho(w_T )\leq 0,
\end{aligned}
\label{teht}
\end{equation}
where $\mathcal{N}$ are stochastic processes adapted to a given filtration $(\mathcal{F}_t)_{t=0}^T$.
 %and $\rho :  L^{0}(\Omega, \mathcal{F}, \mathbb{P}) \to \R$ is a convex risk measure \cite{FollmerSchied}. 
 The variable  $R_t=\sum_{j=1}^JR_t^j\pi_t^j  $ is the return over period $[t-1,t]$ per unit amount of cash invested, and $\pi_t^j$ is the proportion of wealth invested each of the $J$ assets.

%on the filtered probability space $(\Omega, \mathcal{F}, (\mathcal{F})_{t=0}^T,P )$.

Randomness in asset returns gives rise to financial risk, which plays a crucial role in the asset-liability management of a pension plan. Uncertainty in the liabilities consists of both randomness in the in the index that the benefit is tied to, and the mortality risk, which can be decomposed into systematic and non-systematic mortality components. While the impacts of systematic mortality risk, index risk and financial risk do not depend on the size of the pension portfolio, one would expect the non-systematic risk to decrease as the number of members increases.

\begin{remark}['Best estimate']
In the risk-neutral case where $\rho(X)=E[X]$, it can be shown (see Appendix B) that the required initial wealth is
\[
w_0=\frac{\sum_{t=1}^TE(\Pi_{s=t+1}^TR_sc_t)}{E(\Pi_{s=1}^T R_s)}.
\]
%This corresponds with the entropic risk measure when $\gamma=0$ (onko...?).
In the special case where $R_t$ is independent of both its past values and liabilities $c_t$, we obtain the actuarial 'best estimate'
\[
w_0=\sum_{t=1}^T\frac{\bar{c_t}}{\Pi_{s=1}^tR_s},
\]
where $\bar{c_t}=E(c_t)$.% and $\bar{R_s}=\Pi_{s=1}^TE(R_s)$. 
This is the valuation method used in Solvency II.
\end{remark}

The method of valuation given by (\ref{teht}) differs from the actuarial 'best estimate' and the case of \cite{FollmerKnispel} in two aspects. Our setting takes into consideration not only systematic mortality risk but other systematic risk factors that are essential in determining capital requirements in practise, namely investment returns and index values. On the other, the financial market enables the agent to distribute wealth over the time periods. Consequently, the aggregation property of \cite{FollmerKnispel} cannot be applied in this case, and we can observe the effect of nonsystematic risk for various portfolio sizes. The objective of this paper is to study this phenomenon numerically.

\section{Numerical results}

In the following simulation study all members in the pension scheme are females aged 65 at $t=0$, and the term of the scheme is $T=35$. Each member receives a unit benefit per year.
The survival probabilities $p_t$, index $I_t$ and investment returns $R^j_t$ are modelled as a multivariate stochastic process as described in Appendix A. The risk aversion parameter value was set to $\gamma=0.05$.
We generated $N=500000$ scenarios, computed the final wealth $w_T$ in each scenario for a given initial wealth $w_0$, and approximated the expectation in (\ref{eq:entropic}) as a Monte Carlo estimate. The smallest $w_0$ to yield a nonnegative risk for terminal wealths was obtained with a simple line search. The scenarios in the simulation were generated by Latin hypercube sampling. 

Investment returns depend both on the returns on individual assets and the chosen investment strategy.
We consider a simple two-asset fixed proportions investment strategy on bonds and equities.
{\em Fixed proportions} (FP) is a strategy where, in the presence of $J$ assets, the allocation is rebalanced at the beginning of each holding period into set proportions given by a vector $\pi\in\reals^J$, the components of which sum up to one. 
%Parameter of the exponential utility function is $\gamma=0.1$. 
In our example we consider two fixed proportions strategies, namely
 \[
 \pi^S=[\pi_{bond}, \pi_{stock}]=[0.75, 0.25]
 \] 
 and
  \[
 \pi^R=[\pi_{bond}, \pi_{stock}]=[0.5, 0.5]
 \] 
In the first, 'safer' strategy $\pi^S$ a 75\% weight is placed on bonds and a 25\% weight on equities, whereas in the other, 'riskier' strategy $\pi^R$ the weights are 50\% and 50\%.

\begin{figure}[!ht]
 \begin{center}
 \includegraphics[height=0.6\linewidth, width=1\linewidth, angle=0]{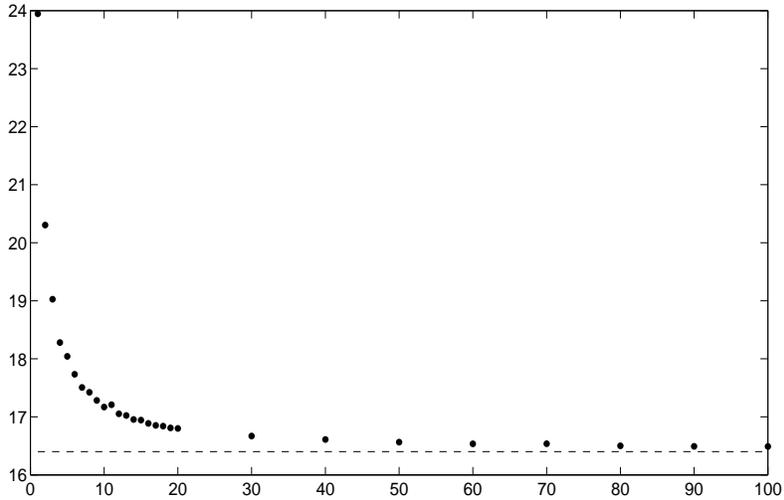} 
 \vspace{-20pt}
 \caption{Initial capital requirement per individual, investment strategy $\pi^S$. Dotted line indicates the level of initial capital required in the presence of systematic risk only.}
\label{fig:w0_0p05}
 \end{center}
 \end{figure}

\begin{figure}[!ht]
 \begin{center}
 \includegraphics[height=0.6\linewidth, width=1\linewidth, angle=0]{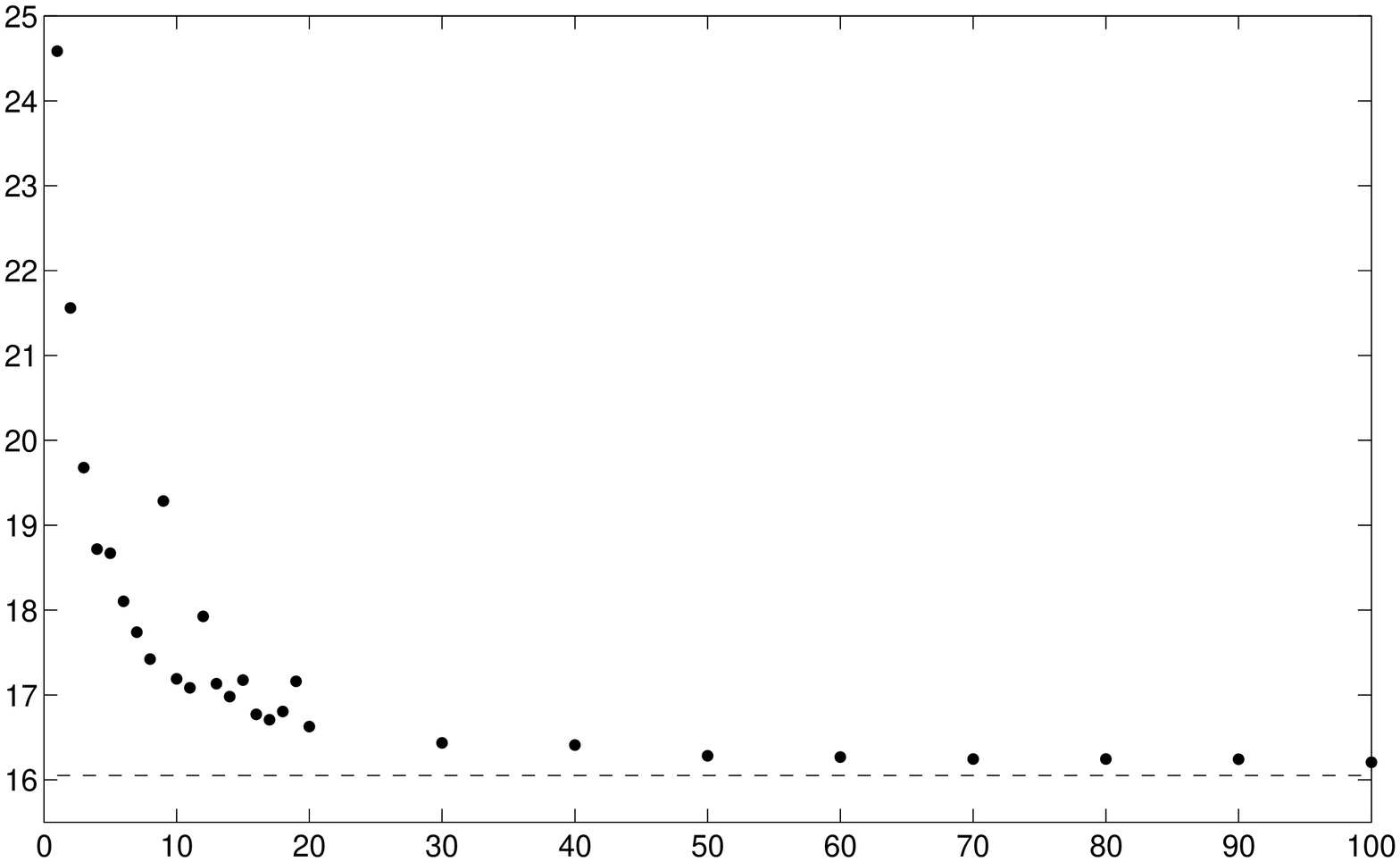} 
 \vspace{-20pt}
 \caption{Initial capital requirement per individual, investment strategy $\pi^R$. Dotted line indicates the level of initial capital required in the presence of systematic risk only. }
\label{fig:w0_0p1}
 \end{center}
 \end{figure}

Figures~\ref{fig:w0_0p05} and~\ref{fig:w0_0p1} plot the initial capitals per individual for various numbers of participants $E_0$ for each strategy, and the two different investment strategies $\pi^S$ and $\pi^R$. The dotted line indicates the level of initial capital required in the presence of systematic risk only, that is when the numbers of survivors are not sampled from binomial distribution but approximated by their expectation as described in Section 2. 
%The solid line marks the actuarial best estimate. 
Initially the required capital drops sharply. With a few dozen members, the effect of nonsystematic risk on the initial capital is already comparatively small. 
%We also note that the higher the risk aversion, the larger is the required additional initial capital per individual, compared with the case of purely nonsystematic risk.

Levels of initial capital required in the risk-neutral case and the actuarial 'best estimate', along with capital required in the presence of systematic risk only for both investment strategies, are presented in Table~\ref{tab:w0_s}. We see that for the risk-neutral case the required capital is slightly smaller than for the actuarial 'best estimate'. This difference arises from the fact that the risk-neutral risk measure takes into account dependencies in asset returns and liabilities. 

%For the purely risk-neutral case of $\rho=E[X]$ the capital requirement amounted to 15.17, whereas the resulting capital for the actuarial best practice was 15.23. 
%The results for $\gamma=0.1$ and $\gamma=0.05$ for the entropic risk measure were 16.69 and 15.87, respectively. 

%
%\begin{table}[!ht]
%\caption{Nonsystematic risk, initial capital}
%\label{tab:w0_ns}
%\footnotesize
%\begin{center}
%\begin{tabular}{|c|ccccccc|}
%\hline 
%   %    & \multicolumn{12}{c|}{}  \\ 
%  $E_0$     & 1&     5 &     10 &   20 & 50 & 100 & $E_0 \to \infty$  \\
%\hline 
%%Female \\
%$\gamma$=0.05 &  19.67 &16.82 & 16.36 & 16.11&  15.97& 15.91& 15.87 \\
%\hline
%$\gamma$=0.1 &  24.98 & 19.76& 17.95  & 17.36& 16.97 &  16.84&16.69   \\
%\hline
%\end{tabular}\\
%\end{center}
%%{\itshape{Notes.}} 
%%Jotain
%\end{table}

\begin{table}[!ht]
\caption{Systematic risk, initial capital}
\label{tab:w0_s}
\footnotesize
\begin{center}
\begin{tabular}{|c|c|c|c|}
\hline 
 Risk measure  & Entropic, $\gamma=0.05$ &   $\rho=E[X]$  & Actuarial 'best estimate' \\
\hline
\hline
 $\pi^S$ & 16.40 & 15.45 & 15.50   \\
 \hline
  $\pi^R$ & 16.05 & 14.09 & 14.12 \\
\hline
\end{tabular}\\
\end{center}
\end{table}

The above illustrations adopt a common assumption that the scheme consists of homogeneous members in terms of age, sex and amount of pension payments. While this setting serves to demonstrate the effect decreasing non-systematic mortality risk, in practice pension plans usually consist of non-homogenous members. 

In order to demonstrate how deviating from the homogeneity assumption may affect the results, we consider a setting where all members in the pension scheme still are females aged 65 at $t=0$, but receive different amounts of pension payment. In our example, 20\% of the initial members of the scheme are entitled to an annual benefit of two units, while 80\% of participants receive the unit benefit, as above. Donnelly \cite{Donnelly} studied the same phenomenon by investigating the effect of including an 'executive section' in the pension plan. We observe that the decrease in non-systematic risk seems to follow a pattern that resembles those associated with homogeneous pension schemes, but the convergence appears to be somewhat slower. However, closer study of this phenomenon falls outside the scope of this work.

\begin{figure}[!ht]
 \begin{center}
 \includegraphics[height=0.6\linewidth, width=1\linewidth, angle=0]{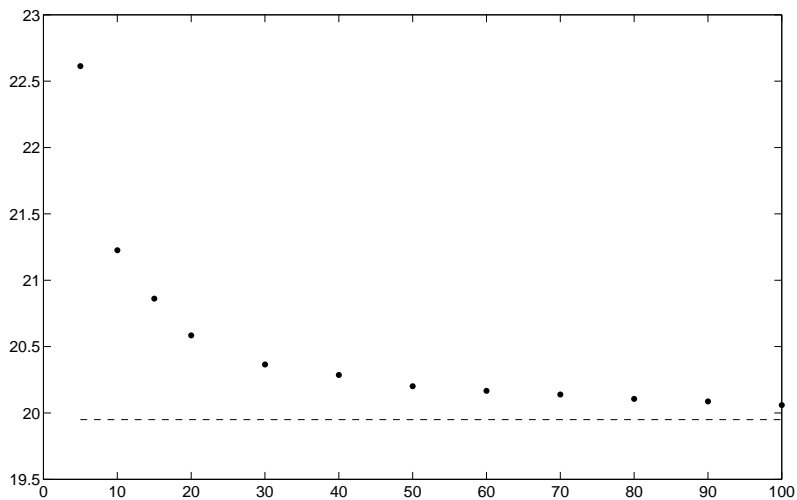} 
 \vspace{-20pt}
 \caption{Initial capital requirement per individual, investment strategy $\pi^S$, non-homogeneous scheme. Dotted line indicates the level of initial capital required in the presence of systematic risk only.}
\label{fig:w0_prem}
 \end{center}
 \end{figure}

\section{Conclusions}

We studied the effects of non-systematic and systematic mortality risks on the required initial capital for a pension plan, in the presence of financial risks. We computed the required initial capital per person for varying numbers of members in the pension scheme. Our main finding was that for pension plans with few members the impact of pooling on the capital requirement per capita is strong, and non-systematic risk is offset rapidly in pension schemes as the number of members increases. Systematic mortality risk, on the other hand, is a significant source of risk is a pension portfolio. 

A possible avenue for future research is closer investigation of how non-homogeneity of members in a pension scheme affects the role and proportion of non-systematic risk. This may include participants that differ in age, gender, or benefit amount.

\appendix

\section{Modelling the systematic risk factors}

In the following we describe a stochastic model for the risk factors affecting the returns and claims of problem \eqref{teht}.

\subsection{Mortality risk factors}

As in \cite{AroPennanen}, we model the {\em survival probabilities} $p_{x,t}$ with the formula
\begin{equation}\label{eq:logit}
p_{x,t} = \frac{\exp\left(\sum_{i=1}^n v^i_t\phi^i(x)\right)}{1+\exp(\sum_{i=1}^n v^i_t\phi^i(x))},
\end{equation}
where $\phi^i$ are user-defined {\em basis functions} and $v^i_t$ are stochastic {\em risk factors} that may vary over time. In other words, the yearly {\em logistic} survival probability of an $x$ year-old is given by
\begin{equation}\label{eq:logit2}
\logit p_{x,t} := \ln \Big(\frac{p_{x,t}}{1-p_{x,t}} \Big) =\sum_{i=1}^nv^i_t\phi^i(x).
\end{equation}
The logistic transformation implies that the probabilities $p_{x,t}$ and $q_{x,t}=1-p_{x,t}$ remain in the interval $(0,1)$. 

By an appropriate choice of the functions $\phi^i(x)$ one can incorporate certain desired features into the model. 
For example, the basis functions can be chosen so that the survival probabilities $p_{x,t}$ have a regular dependence on the age $x$ as e.g.\ in the classical Gompertz model. 

As in \cite{AroPennanen}, we will use the 
three piecewise linear basis functions given by
\begin{align*}
\phi^1(x) &= 
\begin{cases}
1-\frac{x-18}{32} & \text{for $x\le 50$} \\
0 & \text{for $x\ge 50$},
\end{cases}\\
\phi^2(x) &= 
\begin{cases}
\frac{1}{32}(x-18) & \text{for $x\le 50$} \\
2-\frac{x}{50} & \text{for $x\ge 50$},
\end{cases}\\
\phi^3(x) &= 
\begin{cases}
0 & \text{for $x\le 50$}\\
\frac{x}{50}-1 & \text{for $x\ge 50$}.
\end{cases}
\end{align*}
The linear combination $\sum_{i=1}^3 v^i_t\phi^i(x)$ will then be piecewise linear and continuous as a function of the age $x$. %; see Figure~\ref{fig:threeline}. 
The risk factors $v^i_t$ now represent points on logistic survival probability curve: 
\[
v^1_t=\logit p_{18,t},\ v^2_t=\logit p_{50,t},\ v^3_t=\logit p_{100,t}.
\]
It is to be noted that this is just one possible choice of basis functions. Another set of basis functions  would result in another set of risk factors with different interpretations.
% In particular, the Cairns--Blake--Dowd model as described in \cite[Section~4.2]{CBD2008} corresponds to $\phi^1(x)\equiv 1$ and $\phi^2(x)=x-\bar x$, where $\bar x$ is the mean over all ages. In this case, the parameter $v^1$ describes the general level of mortality, while $v^2$ determines how mortality rates change with age. 
We will use the particular three-parameter model described above mainly because of its simple interpretation. %in the economic context. 

Once the basis functions $\phi^i$ are fixed, the realized values of the corresponding risk factors $v^i_t$ can be easily calculated from historical data using standard max-likelihood estimation. The log-likelihood function can expressed as
\begin{align*}
l_t(v) %&=  \ln\prod_{x\in X}{E_{x,t} \choose D_{x,t} } p_v(x)^{D_{x,t}}(1-p_v(x))^{E_{x,t}-D_{x,t}}\nonumber\\
&= \sum_{x\in X}\left[D_{x,t}\sum_{i} v_i\phi_i(x) - E_{x,t}\ln(1+e^{\sum_{i} v_i\phi_i(x)})\right]+d_t\label{ll}
\end{align*}
where $d_t$ is a constant; see \cite{AroPennanen}. The maximization of $l_t$ is greatly facilitated by the fact that $l_t$ is a convex function of $v$; see~\cite[Proposition~3]{AroPennanen}.

\subsection{Investment returns}
\label{al}

Two asset classes, government bonds and and equities, are considered. 
Return on government bonds is given by the formula 
\[
R^b_{t} = \exp(Y_{t-1}\Delta t-D\Delta Y_{t}),
\]
where $Y_{t}^i$ is the yield to maturity at time $t$ and, $D$ is the duration \cite{hylkio}. 
The total return of the equity is calculated as
\[
R_{t}^s = \frac{S_{t}}{S_{t-1}},
\]
where $S_t$ is the total return index. 

\subsection{Time-series model}

Following  \cite{morfin}, we model the future development of risk factors with the following equations
\begin{align*}
\Delta v^{1}_t  &= a^{11}v^{1}_{t-1}+b^1+\varepsilon^1_t  \\
\Delta v^{2}_t &= b^2+\varepsilon^2_t  \\
\Delta v^{3}_t &= a^{33}v^{3}_{t-1}+a^{34}g_{t-1}+b^3+\varepsilon^3_t \\
\Delta g_t&=b^4+\varepsilon^4_t \\
\Delta y_t &= b^5+a^{55}y_{t-1}+\varepsilon^5_t \\
\Delta s_t &= b^6+\varepsilon^6_t ,\\
\Delta p_t &= b^7+a^{77}p_{t-1}+\varepsilon^7_t \\
\end{align*}
where $g_t$ is the logarithm of the per capita GDP, which is in turn modelled as a random walk with a drift. The government bond is the 1-year US Treasury bill, whose yield is denoted by $Y_t$, and log-yield $y_t=\textnormal{log}(Y_t)$. Its equation is the mean reverting interest rate model of Black and Karasinski \cite{BlackKarasinski}. The stock index is the S\&P total return index $S_t$, and its logarithm is described by a random walk with a drift. We denote the consumer price index (CPI) with $P_t$, and model the difference of its logarithm $p_t=\textnormal{log}(P_{t})-\textnormal{log}(P_{t-1})$ as a mean reverting process. 
The terms $\varepsilon_t$ are i.i.d random variables, describing the random fluctuations in the risk factors.

The above equations can be combined to a multivariate linear stochastic difference equation
\[
\Delta x_t= Ax_{t-1}+b+\varepsilon_t
\]
for $x=[v^{1}_t,v^{2}_t,v^{3}_t,  g_t,y_t,s_t,p_t]$. The terms
$\varepsilon_t^i$ are modelled as Gaussian random variables.

The time series model of stochastic risk factors is calibrated to US female mortality and financial market data.
The equations described above were fitted into annual data from 1953--2007. 
US population data was obtained from Human Mortality Database \footnote{www.mortality.org}
Bond yield and consumer price index data was extracted from Federal Reserve Economic Data (FRED) \footnote{http://research.stlouisfed.org/fred2/series/GS1} \footnote{http://research.stlouisfed.org/fred2/series/CPIAUCSL }, and S\&P total return index data from Aswath Damodaran's home page \footnote{pages.stern.nyu.edu/~adamodar/}.  
 
Matrix $A$ and vector $b$ were estimated from data, with the exception of a few user-defined parameters. The value of $b^5$ was chosen such that the mean reversion level of bond log yield corresponds to a yield of $2.5\%$. Similarly, $b^6$ was set to give an average annual equity return of 6\%, and the mean reversion level of the equation for the consumer price index corresponds with an annual inflation rate of $2\%$. Coefficient matrices A and b and covariance matrix $\Sigma$ are as follows:

\[
A = 
\left[\begin{array}{ccccccc} -0.0302 & 0 & 0 & 0 & 0 & 0 & 0\\
 0 & 0 & 0 & 0 & 0 & 0 & 0\\ 0 & 0 & -0.181 & 0.0831 & 0 & 0 & 0\\
  0 & 0 & 0 & 0 & 0 & 0 & 0\\ 0 & 0 & 0 & 0 & -0.209 & 0 & 0\\
   0 & 0 & 0 & 0 & 0 & 0 & 0\\ 0 & 0 & 0 & 0 & 0 & 0 & -0.192
\end{array}\right]
\]

\[
b=
\left[\begin{array}{ccccccc} 0.243 & 0.0139 & -0.673 & 0.0201 & 0.192 & 0.0583 & 0.038 \end{array}\right]
\]

\[
\Sigma = 
\left[\begin{array}{ccccccc}
0.0027&0.0007&0.0014&-0.0004&0.0002&0.0024&0.0003\\
0.0007&0.0006&0.0004&-0.0000&0.0014&0.0010&0.0001\\
0.0014&0.0004&0.0032&-0.0004&-0.0014&0.0000&0.0003\\
-0.0004&-0.0000&-0.0004&0.0005&0.0031&-0.0004&0.0001\\
0.0002&0.0014&-0.0014&0.0031&0.0947&0.0011&0.0035\\
0.0024&0.0010&0.0000&-0.0004&0.0011&0.0246&-0.0010\\
0.000&0.0001&0.0001&0.0001&0.0002&0.0005&0.0002\\
\end{array}\right]
\]

\section{Remark 1: 'Best estimate'}
We obtain, by recursion, 
\[
w_T=w_0\Pi_{s=1}^TR_s-\sum_{t=1}^T(\Pi_{s=t+1}^TR_s)c_t,
\]
where $\Pi_{s=T+1}^TR_s :=1$. Now $\rho(X)=E(X)$,  and we look for $w_0$ for which $E(w_T)=0.$
%\[
%E(w_T)=w_0E(\Pi_{s=1}^TR_s)-\sum_{t=1}^TE(\Pi_{s=t+1}^TR_sc_t)=0,
%\]
We get
\[
w_0=\frac{\sum_{t=1}^TE(\Pi_{s=t+1}^TR_sc_t)}{E(\Pi_{s=1}^T R_s)}.
\]
If we assume that $R_t$ are independent of the liabilities $c_t$, we obtain 
\[
E(\Pi_{s=t+1}^TR_sc_t)=E(c_t)E(\Pi_{s=t+1}^TR_s),
\]
and, further, if $R_t$ are assumed to be independent of their past values, we arrive at
\[
E(\Pi_{s=t+1}^TR_sc_t)=E(c_t)\Pi_{s=t+1}^TE(R_s), 
\]
which yields
\[
w_0=\sum_{t=1}^T\frac{\bar{c_t}}{\Pi_{s=1}^tR_s},
\]
where $\bar{c_t}=E(c_t)$.% and $\bar{R_s}=\Pi_{s=1}^TE(R_s)$. 

\bibliography{bin}
\bibliographystyle{plain}

\end{document}